\documentclass{PoS}

\usepackage{wrapfig}

\title{The LHC p+Pb run from the nuclear PDF perspective}

\ShortTitle{The LHC p+Pb run from the nuclear PDF perspective}

\author{\speaker{Hannu Paukkunen} \\
        Department of Physics, University of Jyv\"askyl\"a, P.O. Box 35, \\ FI-40014 University of Jyv\"askyl\"a, Finland \\
        Helsinki Institute of Physics, University of Helsinki, P.O. Box 64, FI-00014, Finland\\
        E-mail: \email{hannu.paukkunen@jyu.fi}}

\abstract{The p+Pb and Pb+Pb runs at the LHC have opened a possibility to investigate the validity of collinear
factorization in a clearly higher center-of-mass energy scale than earlier in nuclear collisions. Indeed, some processes 
that have been measured routinely in p+p($\overline {\rm p}$) collisions and utilized for years in free proton 
PDF fits, can now finally be reached also in the nuclear case. Such new data are expected to provide conclusive 
answers concerning the universality of the nuclear PDFs. In this talk, I will contrast some of the first p+Pb and
Pb+Pb measurements to the predictions based on the nuclear PDFs.}

\FullConference{XXII. International Workshop on Deep-Inelastic Scattering and Related Subjects,\\
		28 April - 2 May 2014\\
		Warsaw, Poland}

\begin{document}

\section{Introduction}

The LHC proton+lead (p+Pb) run with $\sqrt{s}=5.02 \, {\rm TeV}$ and lead+lead (Pb+Pb) runs with $\sqrt{s}=2.76 \, {\rm TeV}$
center-of-mass energy have brought the nuclear collisions to a completely new energy realm. Indeed, before the
LHC-era the record energy in nuclear collisions was only $\sqrt{s}=0.2 \, {\rm TeV}$ reached at RHIC-BNL. This increase
in center-of-mass energy by more than an order of magnitude has given a possibility to look processes like on-shell
heavy gauge boson production and high-transverse-momentum jets that are difficult (or impossible) to measure at RHIC. In this
talk I will discuss some of the recent LHC data from nuclear collisions, to what extent they are consistent
with the collinear-factorization-based expectations
$$
 \sigma^{A+B \rightarrow \mathcal{O}} = \sum_{i,j} 
  \mathop{\mathop{\underbrace{f_i^A(\mu_{\rm fact}^2)}}_
 {\rm nuclear \,\, PDFs, \,\, obey}}_
 {\rm the \,\, usual \,\, DGLAP}
 \otimes \,\,\,\, 
  \mathop{\mathop{\underbrace{\hat \sigma^{i+j \rightarrow \mathcal{O}}(\mu_{\rm fact}^2,\mu_{\rm ren}^2)}}_
 {\rm usual \,\, pQCD}}_
 {\rm coefficient \\ \,\, functions} 
 \,\,\,\,  \otimes \mathop{\mathop{\underbrace{f_j^B(\mu_{\rm fact}^2)\, ,}}_
 {\rm nuclear \,\, PDFs, \,\, obey}}_
 {\rm the \,\, usual \,\, DGLAP}
$$
whether there is evidence for nuclear modifications in parton distribution functions (PDFs) $f_i^A$, and are they 
consistent with the pre-LHC extractions \cite{Paukkunen:2014nqa}.

\section{The CMS dijet measurements}

\begin{figure}[th!]
\centering
\includegraphics[width=0.45\textwidth]{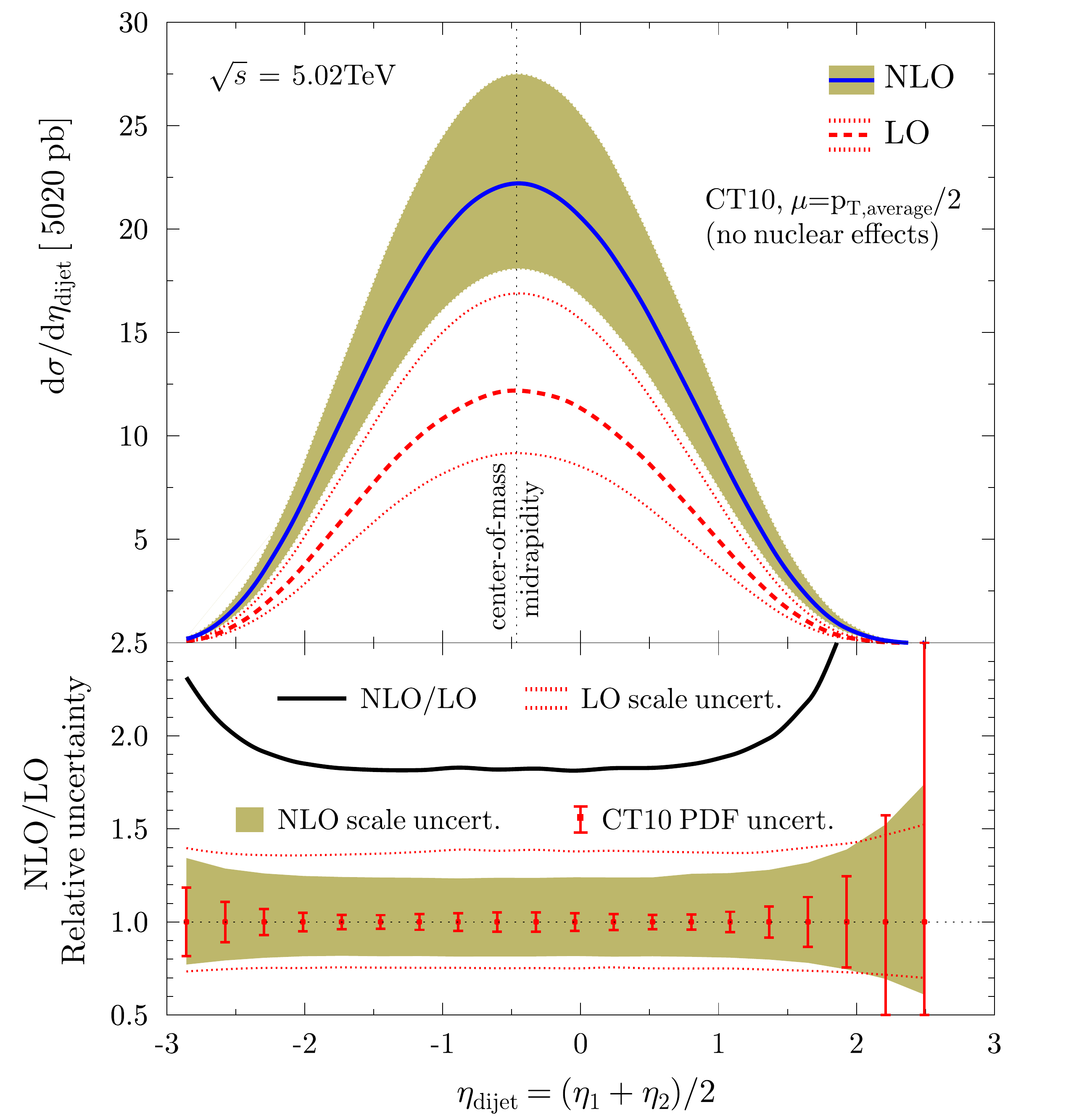}
\includegraphics[width=0.45\textwidth]{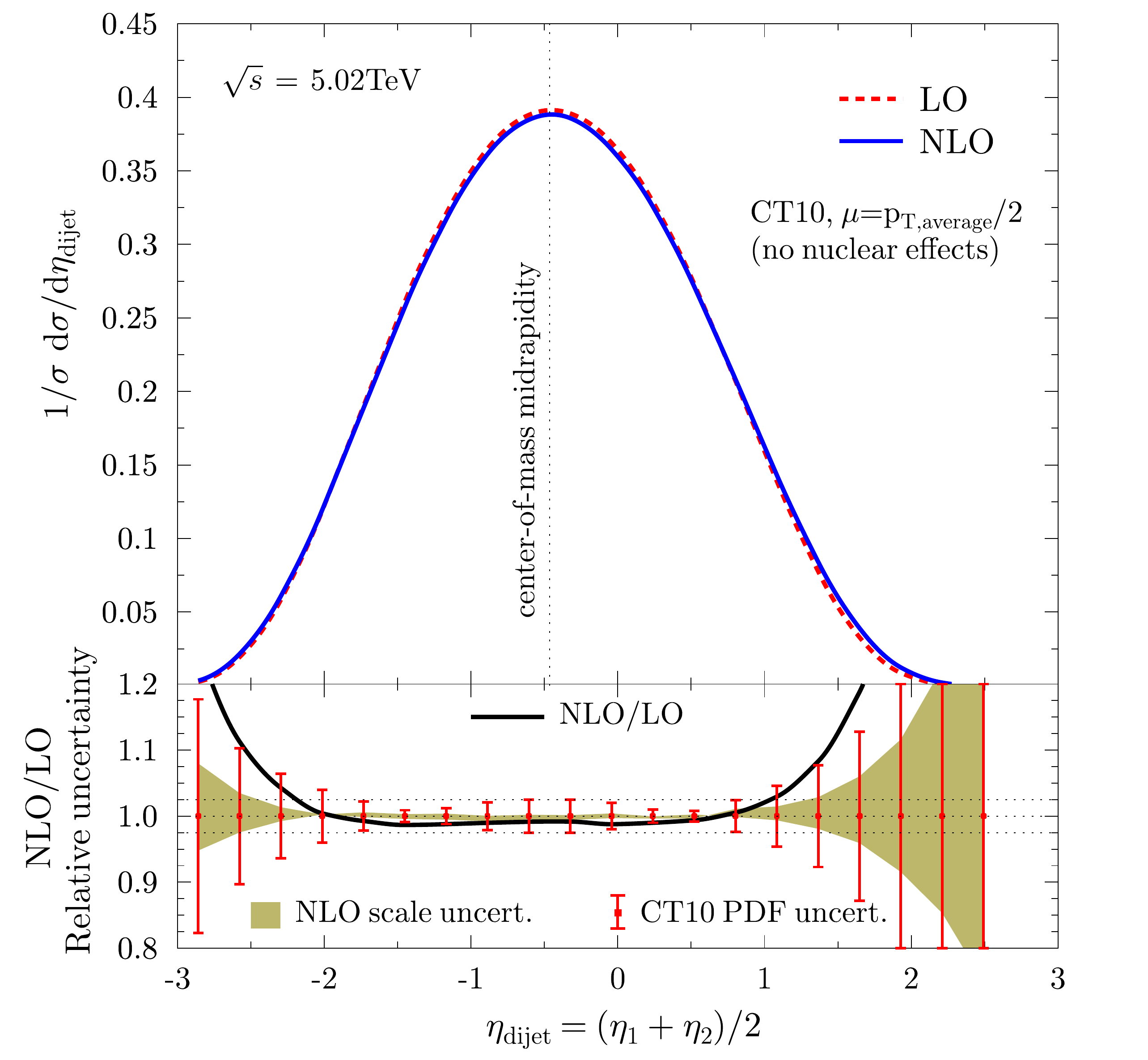}
\caption{Predictions for the absolute dijet spectrum (left) and the
same distribution normalized by the rapidity-integrated cross section.
Figures from \cite{Eskola:2013aya}.} 
\label{fig:spectra}
\end{figure}

One of the very first data that came out soon after the 2013 p+Pb run was the dijet measurement of the CMS 
collaboration \cite{Chatrchyan:2014hqa}. The data was presented in bins of dijet ``pseudorapidity'' 
($\eta_{\rm leading \,\, jet}$ and $\eta_{\rm sub-leading \,\, jet}$ are the pseudorapidities of the leading and subleading jet)
\begin{equation}
 \eta_{\rm dijet} \equiv \frac{1}{2} \left( \eta_{\rm leading \,\, jet} + \eta_{\rm sub-leading \,\, jet} \right), 
\end{equation}
with $p_{T, \rm leading \,\, jet} > 120 \, {\rm GeV}$ and $p_{T, \rm leading \,\, jet} > 30 \, {\rm GeV}$. At
the time of the data taking the proton beam was set to $E_{\rm p}=4 \, {\rm TeV}$, and the lead-ions
circulated with $E_{\rm Pb}= (82/208) \times 4 \, {\rm TeV} \approx 1.58 \, {\rm TeV}$ correspondingly. 
For these unequal energies of the colliding nucleons the center-of-mass midrapidity shifted about half
a unit in the  laboratory frame (indicated in the following figures).

The perturbative QCD calculations \cite{Eskola:2013aya} for the absolute dijet spectrum at leading order and next-to-leading
order (NLO) in strong coupling $\alpha_s$ are shown in Figure~\ref{fig:spectra} (with no nuclear effects, just using CT10NLO \cite{Lai:2010vv}).
The NLO correction is always large, almost a factor of two, and the scale uncertainty is around 20\% or more for
not imposing a lower cut on the dijet invariant mass. All the details of nuclear modifications in PDFs could be easily hidden
under such large uncertainties. However, a bit surprisingly, the \emph{shape} of this distribution around the central rapidities
does not appear to receive large NLO corrections as demonstrated in the right-hand panel of Figure~\ref{fig:spectra} 
where the absolute distribution has been normalized by the total cross section.
\begin{figure}[th!]
\centering
\includegraphics[width=0.45\textwidth]{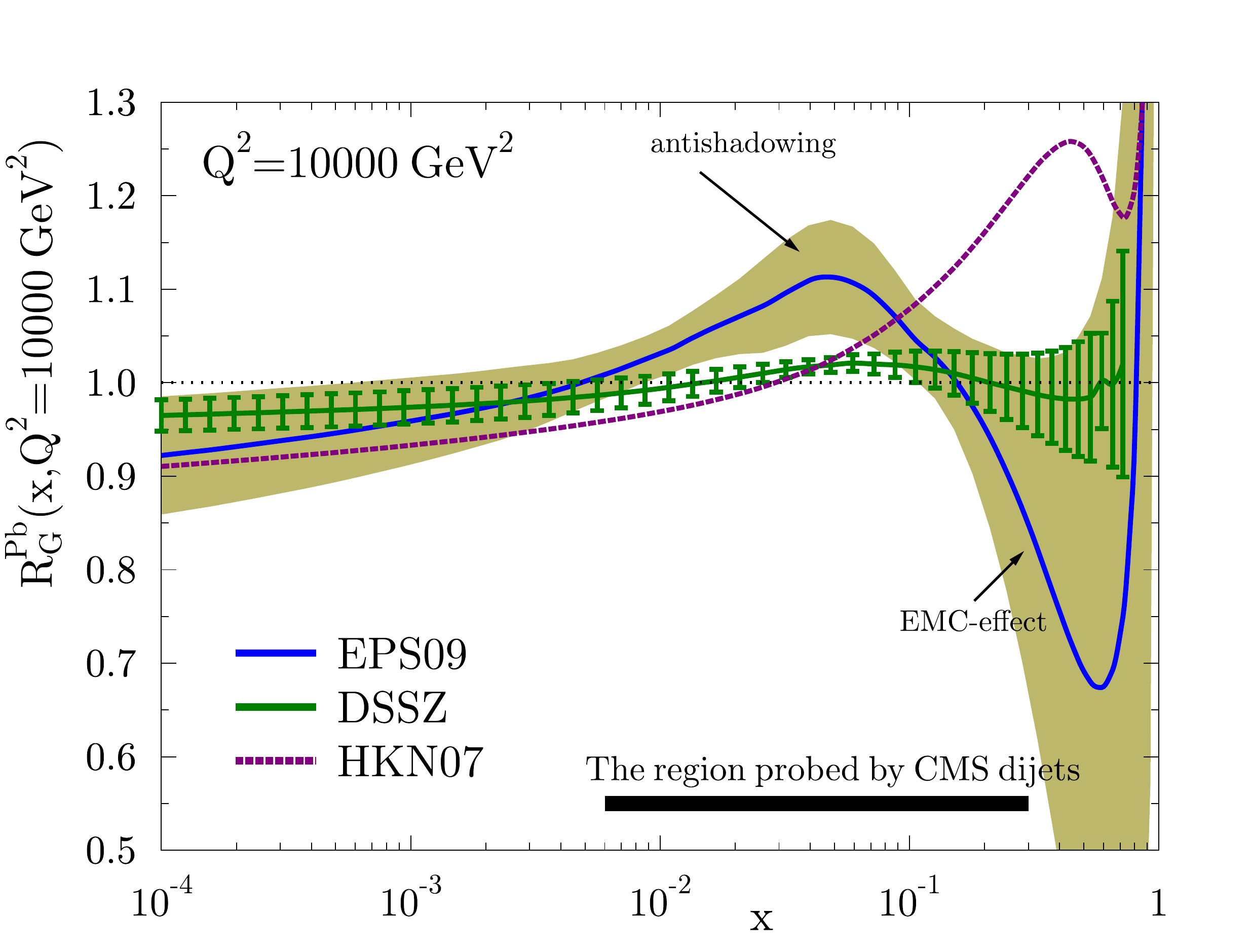}
\includegraphics[width=0.45\textwidth]{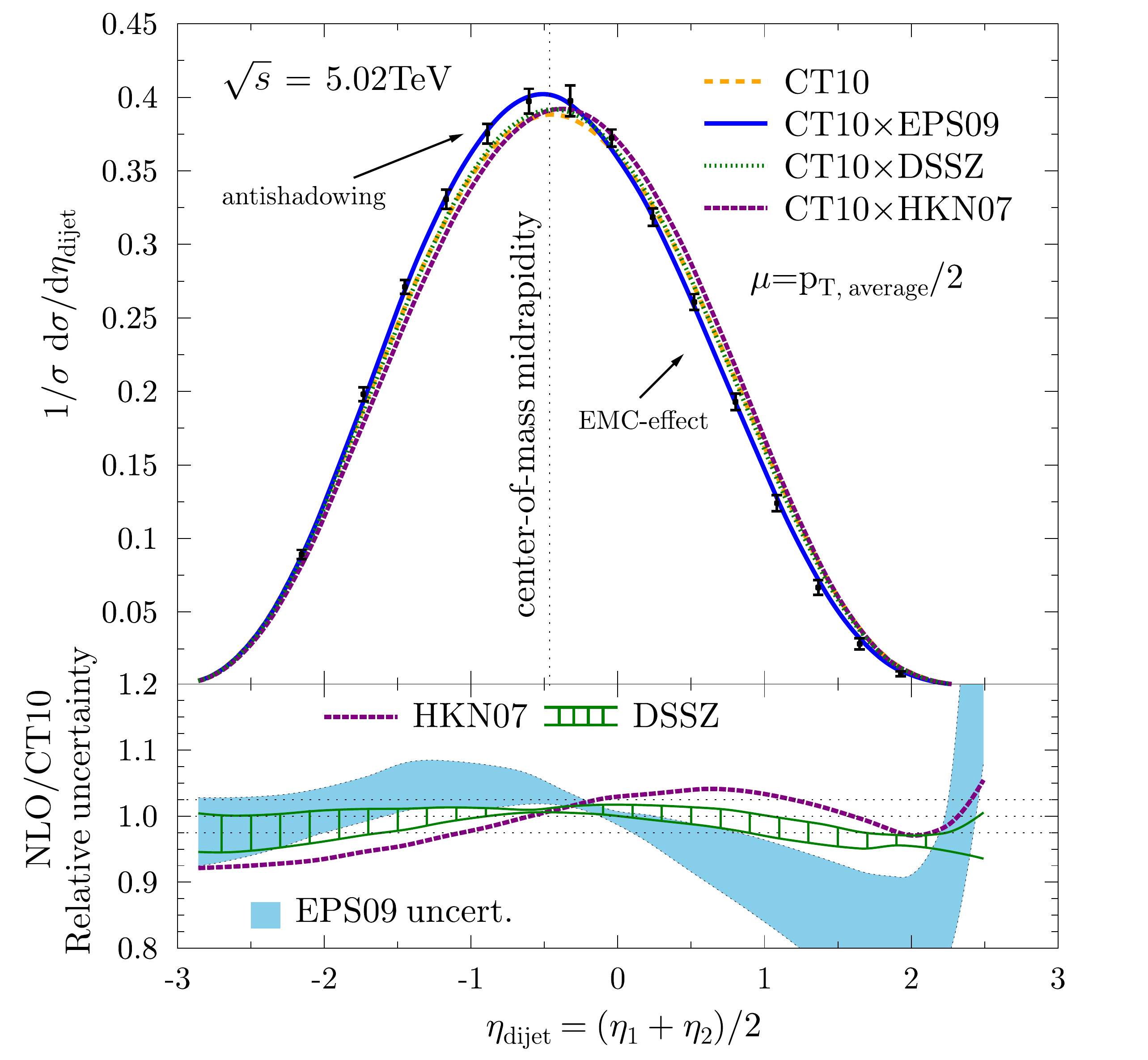}
\caption{Nuclear modifications in gluon PDFs $R_{\rm G}^{\rm Pb}=g^{\rm Pb}(x,Q)/g^{\rm p}(x,Q)$
from different parametrizations and the resulting predictions for the normalized dijet
distributions together with the preliminary CMS data.} 
\label{fig:spectra2}
\end{figure}
In Figure~\ref{fig:spectra2}, the preliminary CMS data \cite{Chatrchyan:2014hqa} for this normalized distribution is compared to the calculations
with no nuclear effects (only CT10NLO), and with various nuclear PDFs (nPDFs), HKN07 \cite{Hirai:2007sx},
DSSZ \cite{deFlorian:2011fp} and EPS09 \cite{Eskola:2009uj}. From these, only EPS09 is capable to systematically 
reproduce the the data. The reason can be tracked down to the gluon ``antishadowing'' and ``EMC-effect'' indicated
as well in Figure~\ref{fig:spectra2} which are absent in the other parametrizations.\footnote{The preliminary nCTEQ
parametrization has also similar effects. See Ref.~\cite{Kovarik:2013sya} and the talk by A.~Kusina, DIS2014.}
As these effects in EPS09 were
inferred from RHIC inclusive pion production data at considerably lower center-of-mass energy and much lower transverse
momentum ($p_T < 10 \, {\rm GeV})$ as well, the agreement here lends support for the conjecture of the nPDF universality.

\section{The dilemma of inclusive charged-hadron production}

\begin{wrapfigure}{r}{0.60\textwidth}
\vspace{-1cm}
\centerline{\includegraphics[width=0.5\textwidth]{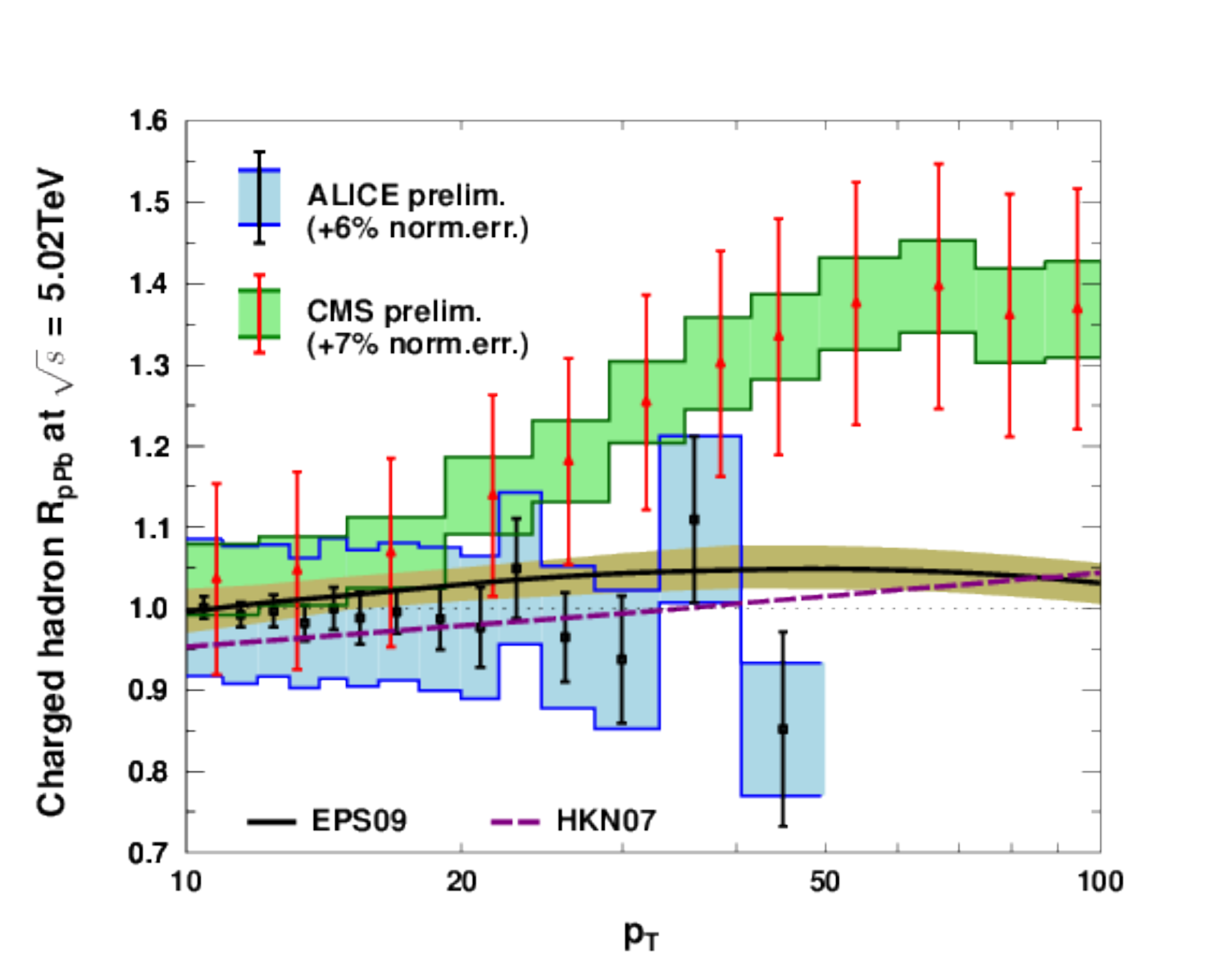}}
\caption[]{The nuclear modification of inclusive charged-hadron production at midrapidity as
measured by ALICE \cite{Abelev:2014dsa} and CMS \cite{CMS:2013cka} compared to the nPDF predictions
(EPS09 and HKN07).}
\label{fig:RpA}
\end{wrapfigure}

An issue that has recently caused some confusion is the large enhancement seen in the nuclear modification
factor $R_{\rm pPb} \equiv d\sigma_{\rm p+Pb}/d\sigma_{\rm p+p}$ for inclusive high-$p_T$ charged-hadron 
production reported by the CMS collaboration \cite{CMS:2013cka}, and shown here in Figure~\ref{fig:RpA}.
An enhancement as large as this came completely unexpected and is far too large to stem from nuclear
modifications in PDFs. However, the same observable measured by ALICE \cite{Abelev:2014dsa} shows
no sign of such enhancement although the $p_T$ reach is more restricted than that of CMS. It should be
noted that there is no baseline p+p measurement ($d\sigma_{\rm p+p}$) at $\sqrt{s}=5.02 \, {\rm TeV}$
but it has to be constructed by other means to form the nuclear modification factor $R_{\rm pPb}$. This could partly
contribute to the significant difference between the CMS and ALICE results.

\begin{figure}[th!]
\centering
\includegraphics[width=0.30\textwidth]{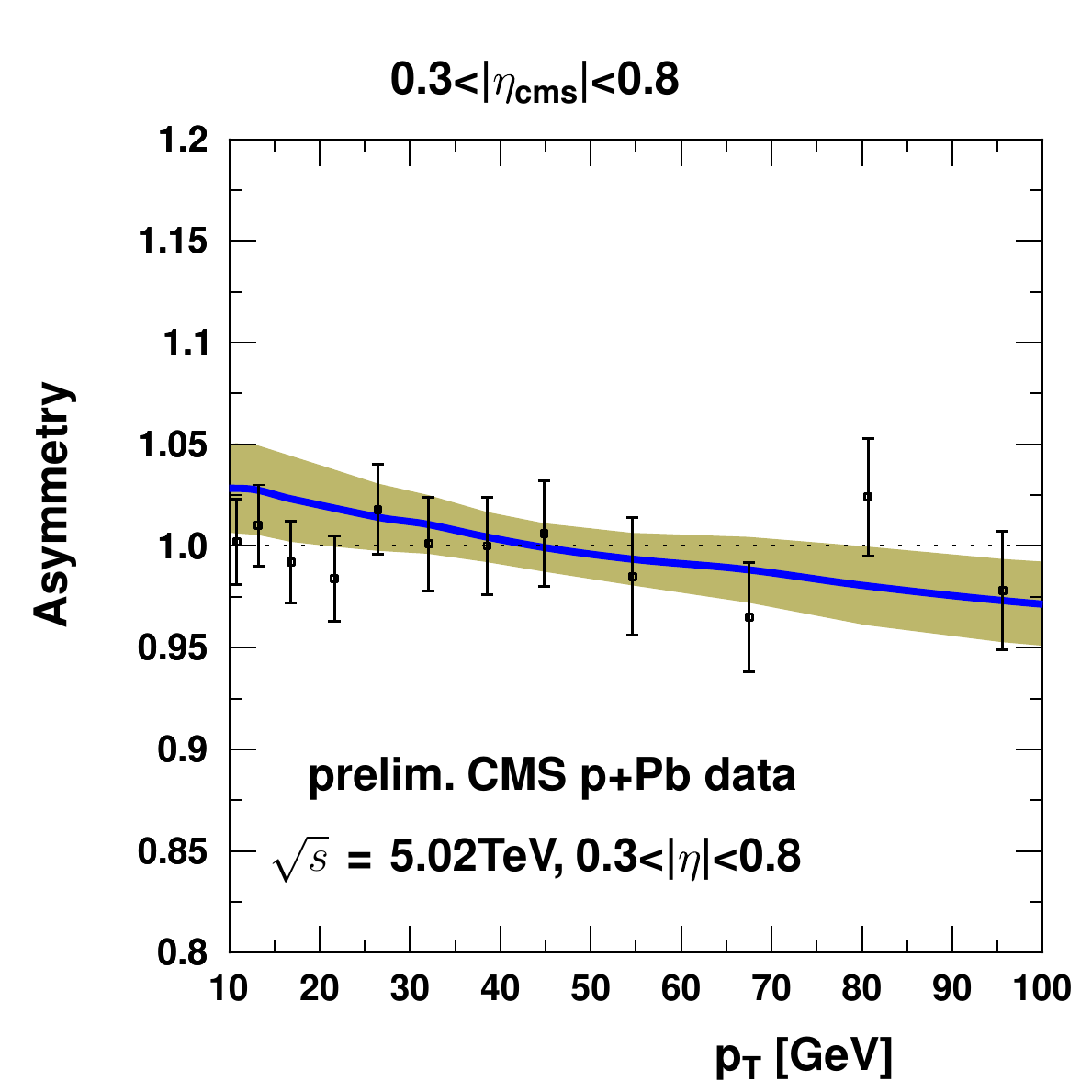}
\includegraphics[width=0.30\textwidth]{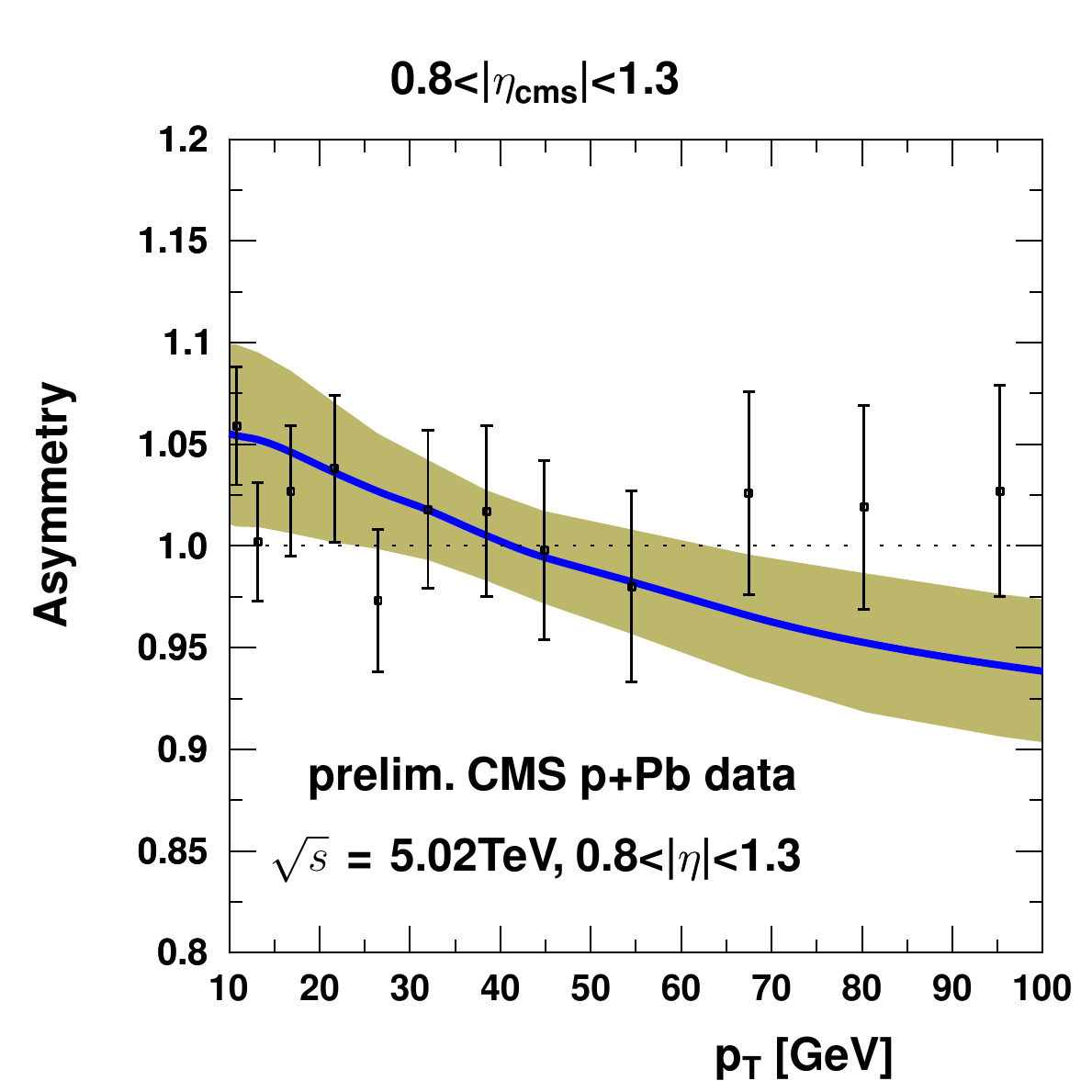}
\includegraphics[width=0.30\textwidth]{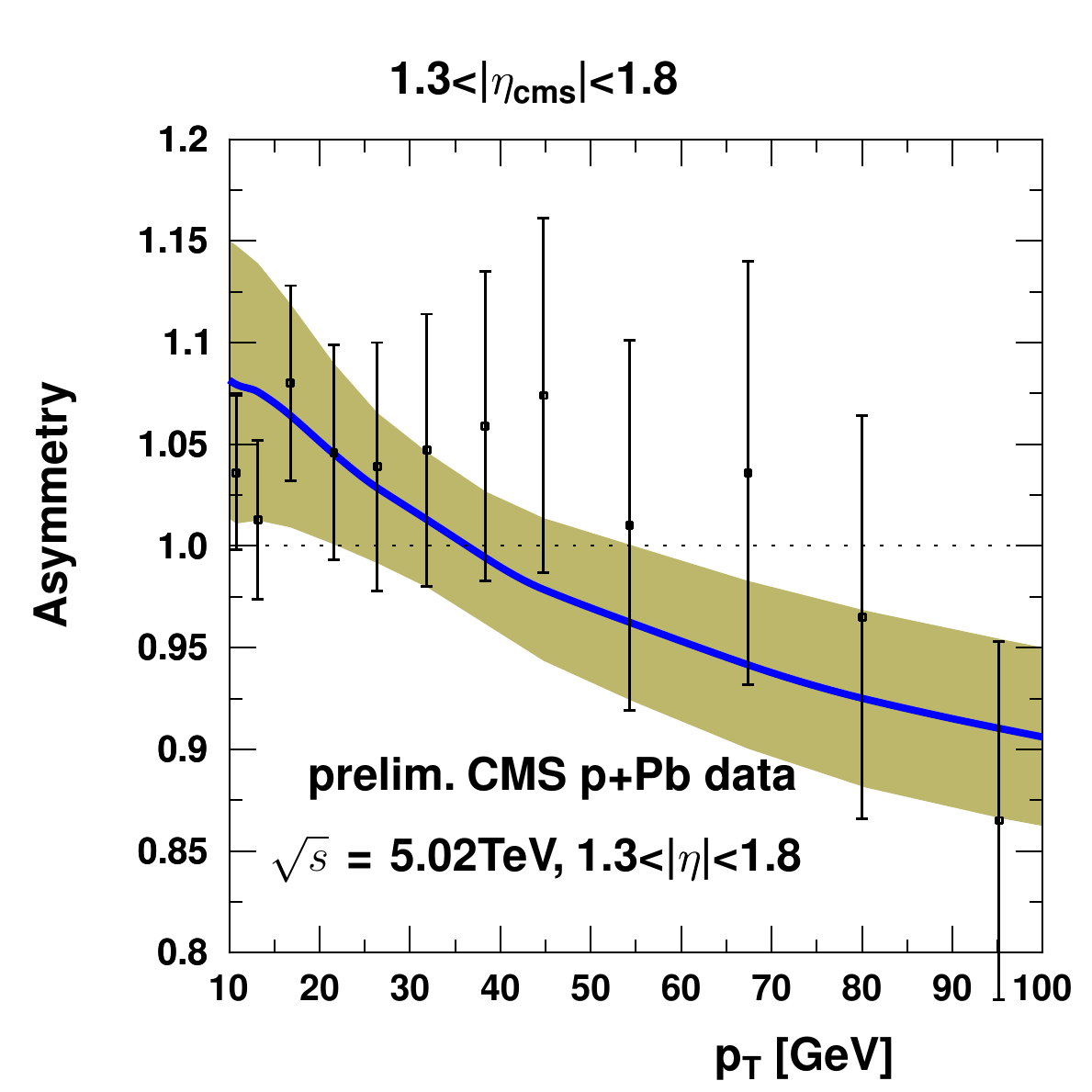}
\caption{Charged-hadron forward-to-backward asymmetry as measured by
CMS compared to the EPS09 predictions. Data points taken from \cite{CMS:2013cka}.} 
\label{fig:asym}
\end{figure}

An observable that sidesteps the need for a p+p baseline is the forward-to-backward asymmetry
$d\sigma(\eta_{\rm cms})/d\sigma(-\eta_{\rm cms})$ which can be measured by CMS for its wide-enough
rapidity acceptance. The results reported in \cite{CMS:2013cka} are shown in Figure~\ref{fig:asym}
and compared to the predictions using EPS09. Interestingly, the sizable enhancement seen in 
Figure~\ref{fig:RpA} appears to be independent of the rapidity interval and these measurements 
are more or less consistent with the EPS09 predictions. 

\section{Heavy gauge-boson production}

\begin{figure}[th!]
\centering
\includegraphics[width=0.45\textwidth]{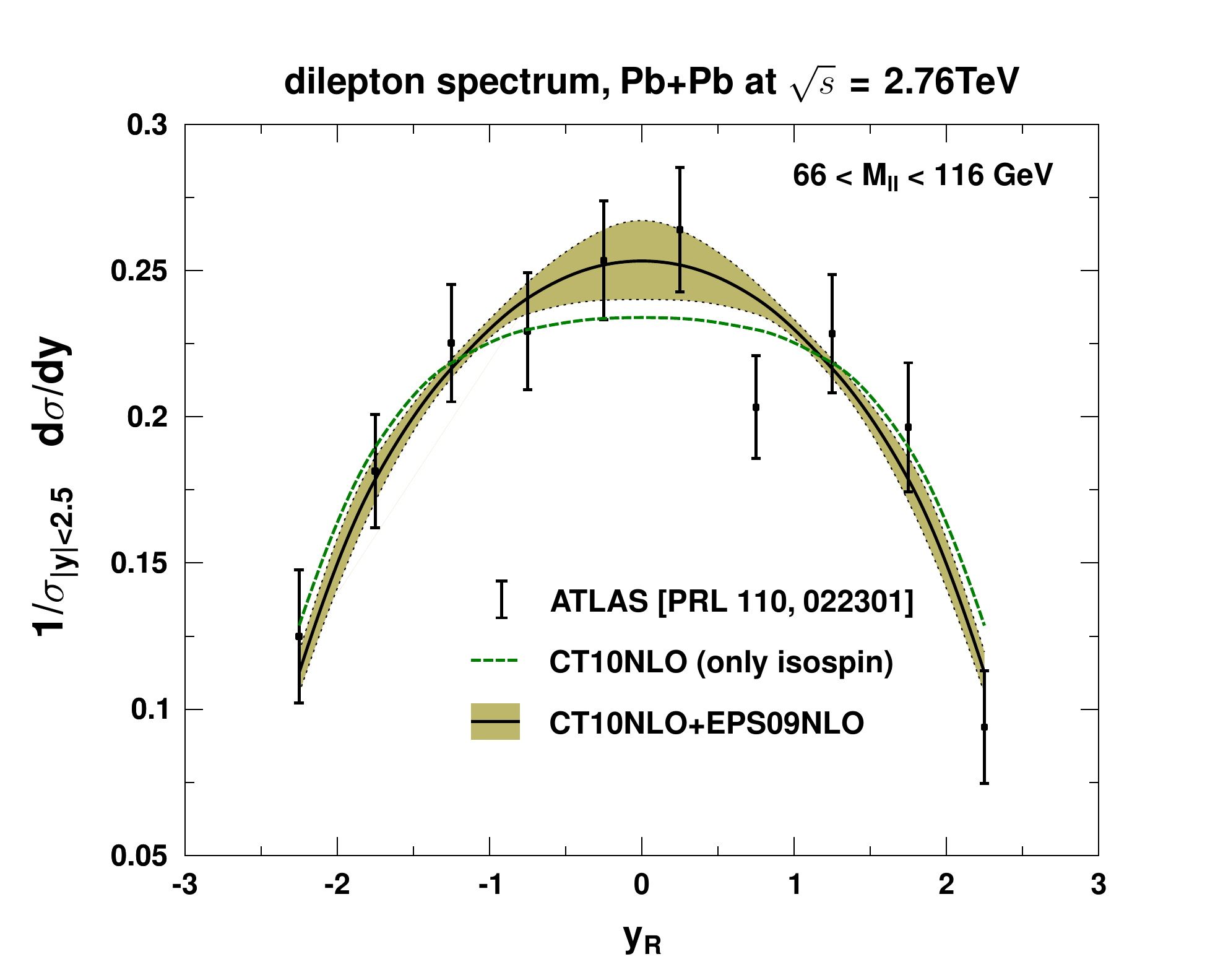}
\includegraphics[width=0.45\textwidth]{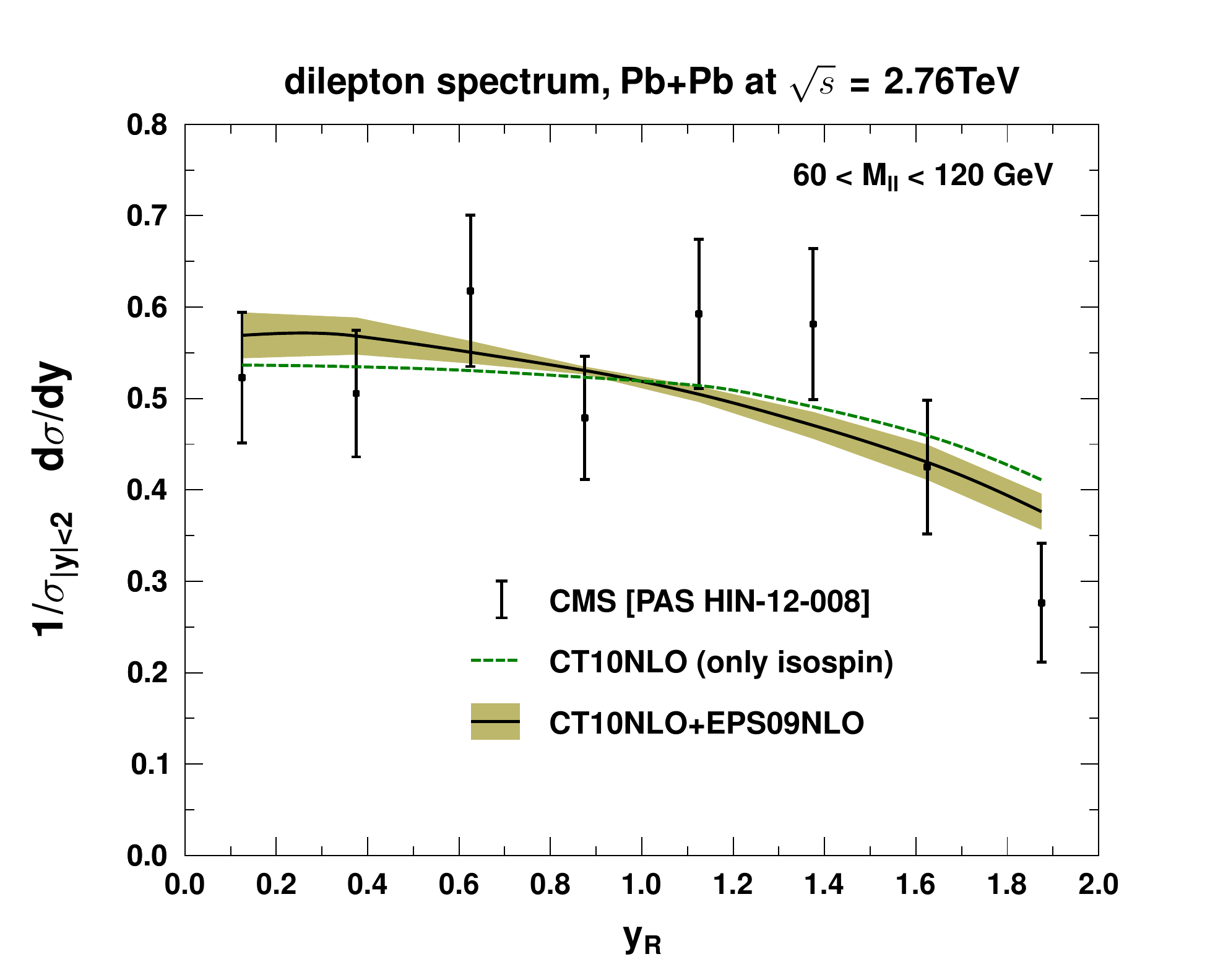}
\caption{The normalized rapidity distribution of Z bosons in Pb+Pb collisions as measured by the ATLAS and 
CMS collaborations compared to the NLO calculations with and without EPS09 nuclear effects
in CT10NLO PDFs. The data points have been obtained from \cite{Aad:2012ew,CMS:2012sba} and
the integrated yields (the normalization factors) have been estimated by summing up the central
data values multiplied by the bin widths.} 
\label{fig:PbPbZ}
\end{figure}

\begin{wrapfigure}{r}{0.55\textwidth}
\vspace{-0.7cm}
\centerline{\includegraphics[width=0.45\textwidth]{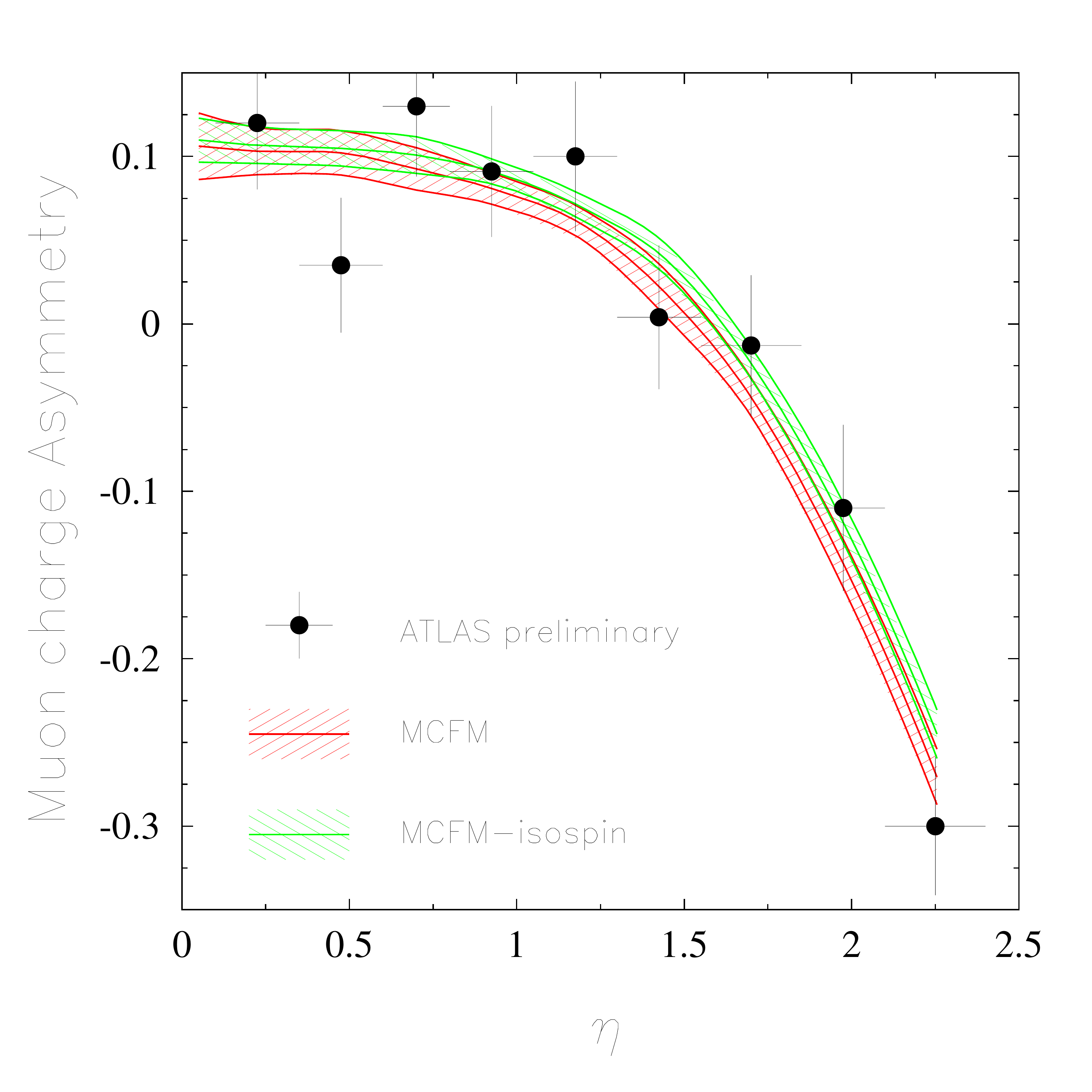}}
\caption[]{The charge asymmetry in Pb+Pb collisions as measured by ATLAS \cite{ATLAS-CONF-2013-106} compared to 
calculations with (red band) and without (green band) EPS09 nuclear effects.
Figure by P.~Zurita.}
\label{fig:PbPbW}
\vspace{-0.7cm}
\end{wrapfigure}

Production of charged leptons from heavy gauge-boson decays probes the quark sector of the nPDFs.
Theoretically well-understood (e.g. small scale uncertainty, known up to next-to-NLO) observables
are the rapidity distributions of opposite-charge di-lepton pairs ($Z$ production) and single charged
leptons ($W^\pm$ production). The first nuclear data for such observables come from the LHC Pb+Pb 
collisions at  $\sqrt{s}=2.76 \, {\rm TeV}$ center-of-mass energy. While the precision of these Pb+Pb data is
not high enough to decide whether we see evidence for the presence of nuclear modifications in PDFs,
 the data for $Z$ boson production (shown in Figure~\ref{fig:PbPbZ})
and those for the charge asymmetry in production of $W^\pm$ bosons (shown in Figure~\ref{fig:PbPbW}) follow 
very nicely the predictions derived from the collinear factorization.

\begin{figure}[th!]
\centering
\includegraphics[width=0.35\textwidth]{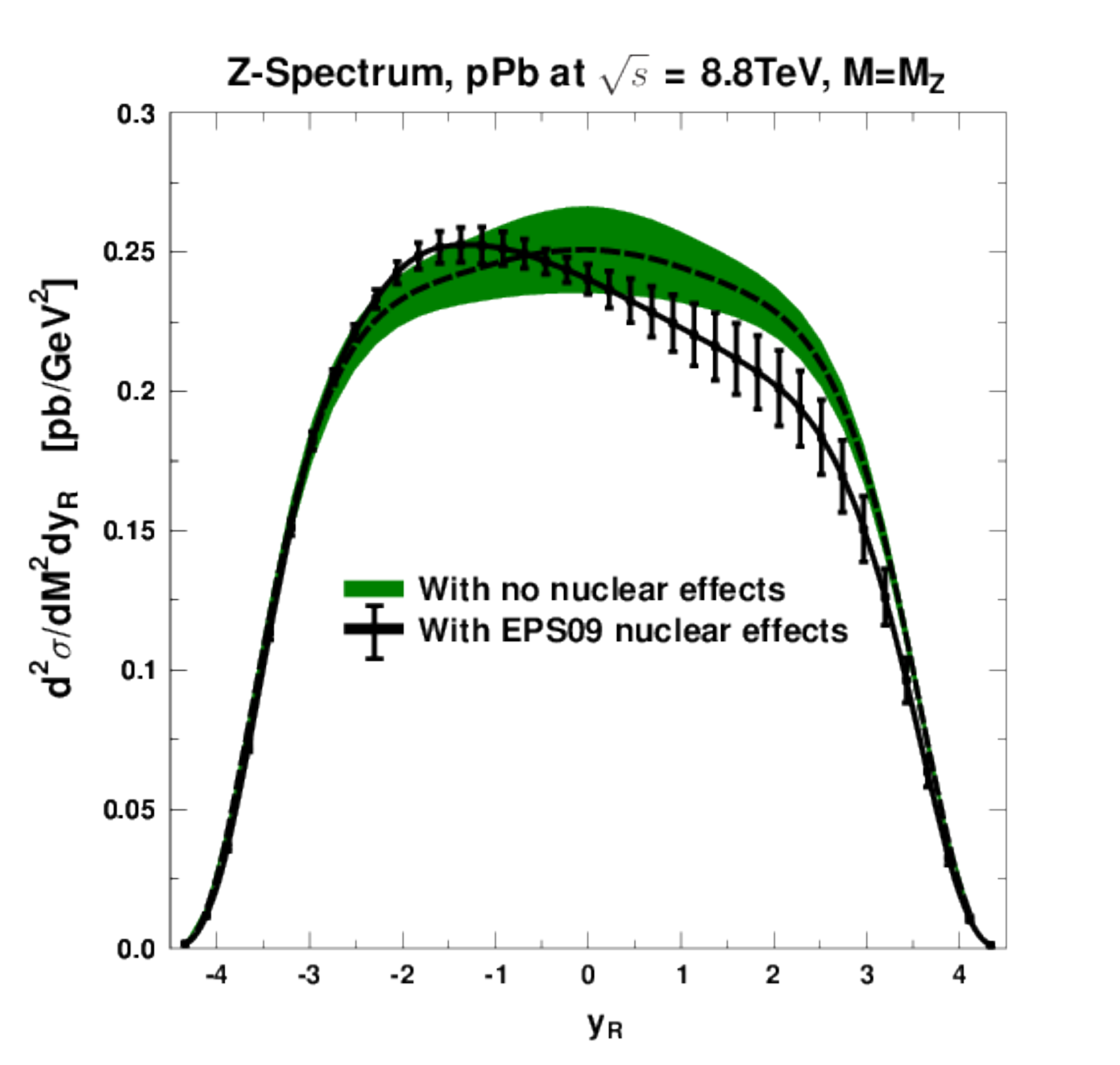}
\includegraphics[width=0.60\textwidth]{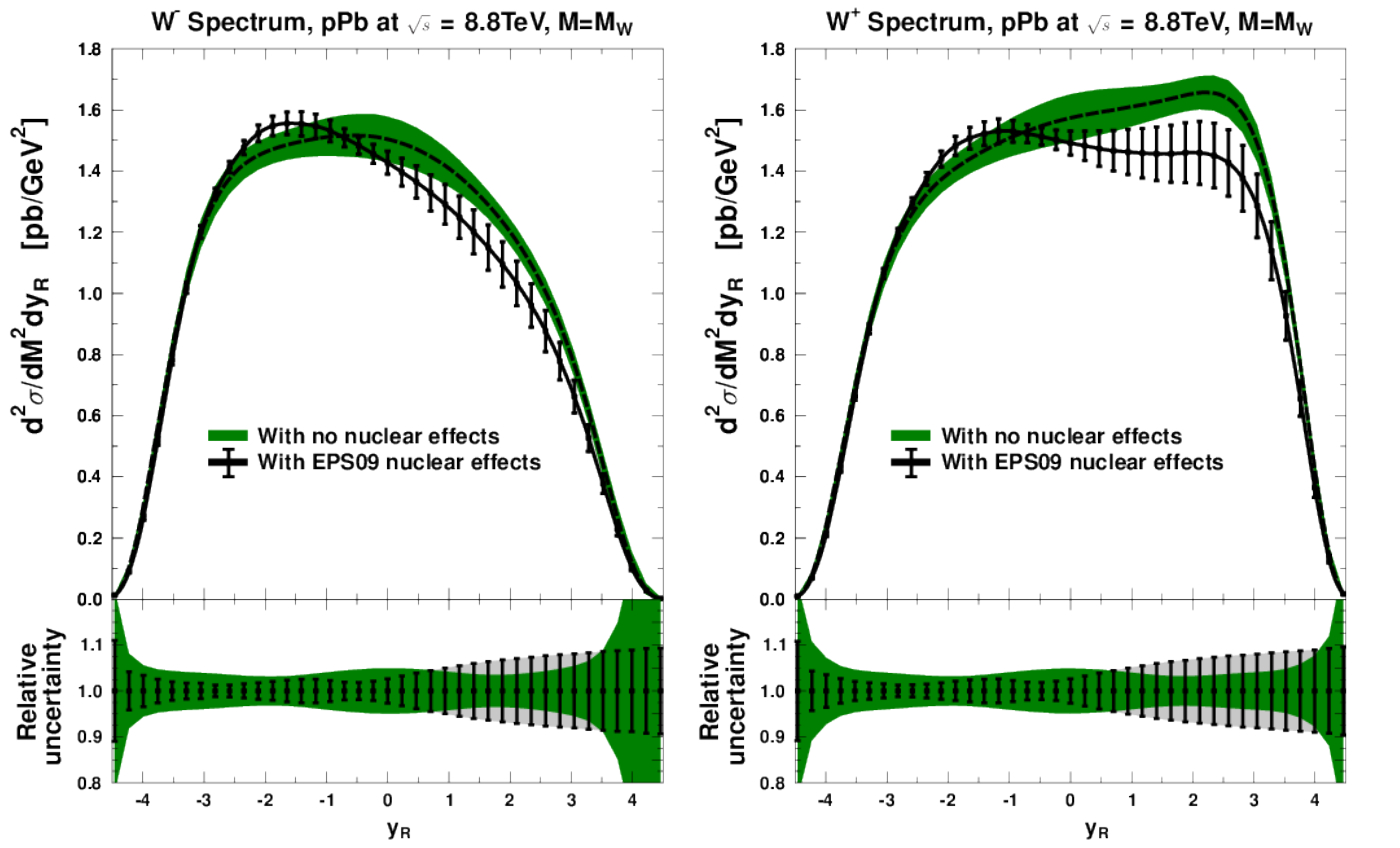}
\caption{Illustrative examples of the predicted $Z$ and $W^\pm$ rapidity distributions 
in p+Pb collisions at the LHC and the expected impact of the nuclear modifications in PDFs.
Figure from \cite{Paukkunen:2010qg}.} 
\label{fig:ZW}
\end{figure}

Similarly to the case of dijets discussed earlier, the rapidity distributions of leptons from $Z$ and $W^\pm$
decays in p+Pb collisions are expected to be different in the proton- and
lead-going directions. Examples illustrating what can be generally expected are plotted in Figure~\ref{fig:ZW}.
As the collected data sample in 2013 p+Pb run is significantly larger than that of the 2011
Pb+Pb run, the precision should consequently be clearly better than the Pb+Pb data shown in Figures~\ref{fig:PbPbZ}
and \ref{fig:PbPbW}. Thus, there is hope that, once available, the p+Pb data for $Z$ and $W^\pm$ production
should provide constraints for the nPDFs.

\vspace{-0.2cm}
\section{Summary}

\vspace{-0.2cm}
In summary, I have presented some of the LHC p+Pb and Pb+Pb results from the
nuclear PDF perspective. Remarkably, the first p+Pb dijet measurement by the CMS
collaboration points towards the existence of gluon antishadowing. If confirmed, this would be in accord
with the inclusive pion data measured at RHIC. The inclusive
minimum-bias charged-hadron production has been measured both by CMS and ALICE and 
there is a clear inconsistency between these two independent measurements. This underscores the importance to carry out
p+Pb measurements in all major LHC experiments to be able to cross-check results. 
The $Z$ and $W^\pm$ production have a high potential to provide novel constraints for the nuclear quarks.
Even in Pb+Pb collisions the collinear factorization appears to work rather well for these 
processes and the expectations with respect to the upcoming p+Pb data can thus be set high.

\vspace{-0.2cm}
\section*{Acknowledgments}
\noindent I acknowledge the financial support from the Academy of Finland, Project No. 133005.

\end{document}